\documentstyle[12pt,aps,epsf]{revtex}

\begin{document}

\begin{center}
{\large\bf Broadband detection of squeezed vacuum:\\
           A spectrum of quantum states}

{G. Breitenbach, 
F. Illuminati$^*$,
S. Schiller, 
and  J. Mlynek}

\date{\today}

{\sl Fakult\"at f\"ur Physik, Universit\"at Konstanz, D-78457 Konstanz,
         Germany}\\
{\sl http://quantum-optics.physik.uni-konstanz.de}\\
{\sl $^*$ Permanent address: Dipartimento di Fisica, Universit\`a  
di Salerno, and INFM, Unit\`a di Salerno, I-84081 Salerno, Italy}

{\sl Phone: +49(7531) 883842, FAX +49(7531) 883072, 
   e-mail:   Gerd.Breitenbach@uni-konstanz.de}

\end{center}

\vskip3mm
We demonstrate the simultaneous quantum state reconstruction of the 
spectral modes of the light field emitted by a continuous wave
degenerate optical parametric amplifier. The scheme is based on  
broadband measurement of the quantum fluctuations of the
electric field quadratures and subsequent Fourier decomposition  
into spectral intervals. Applying the standard reconstruction algorithms
to each bandwidth-limited quantum trajectory, a "spectrum" of density  
matrices and Wigner functions is obtained. The recorded states show a
smooth transition from the squeezed vacuum to a vacuum state.
In the time domain we evaluated the first order correlation function
of the squeezed output field, showing good agreement with theory.

\vskip10mm
The experimental techniques of quantum state reconstruction, first  
applied five years ago,
have opened a new field of research, wherein simple quantum  
mechanical systems 
can be  characterized completely by density matrices and Wigner  
functions
\cite{vog89}.
Our system consists of electromagnetic field modes at optical  
frequencies.
We have previously generated the whole family of squeezed states of 
light using an optical parametric amplifier (OPA)
and reconstructed these states using the method of optical 
homodyne tomography \cite{bre97a}.
The reconstructions presented therein were limited to essentially one
particular pair of modes at frequencies $\omega\pm\Omega$, where  
$\Omega$ is
a radio frequency and $\omega$ the optical frequency. The spectral  
bandwidth
of this mode pair was $\Delta\Omega/2\pi$ = 100 kHz.
Since an OPA pumped below threshold emits a frequency {\sl spectrum}, 
with a bandwidth determined by the cavity linewidth
in the order of several MHz, the output of the OPA
is described more precisely by a 
whole spectrum of quantum states \cite{col85}.
General schemes for multimode reconstruction become quite complicated
already at the two-mode level \cite{ray96}.
So far only one experiment demonstrating 
photon-number correlations in the time domain by measuring the  
photon statistics
via dual-pulse phase-averaged
homodyne detection has been carried out \cite{mca97}.
In this work we use a simple measurement scheme to record
simultaneously the complete statistical information about 
the quantum  states 
of all mode-pairs emitted by the OPA,
disregarding correlations between the different mode-pairs. 
This enables us 
to reconstruct the Wigner functions and density matrices 
corresponding to a large set of mode pairs.

\vskip3mm

\begin{figure}
\epsfxsize=15cm
\epsfbox[210 152 640 438]{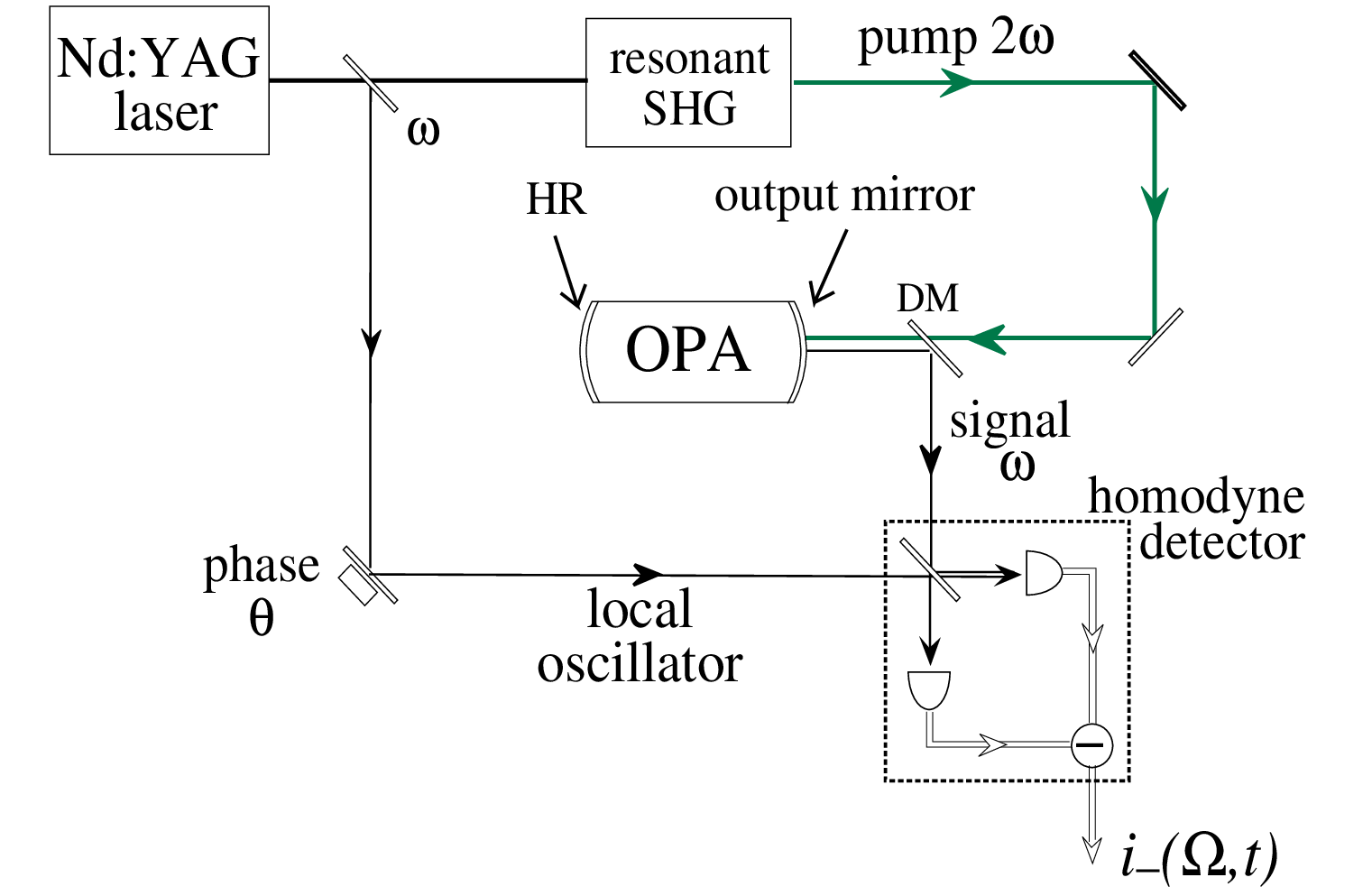}
\caption{Experimental scheme for measuring the quantum states of  
the OPA.
\label{figmult1}}
\end{figure}

The experimental setup is shown in Fig.\ref{figmult1}.  Center  
piece of the experiment
is a monolithic standing-wave lithium-niobate OPA, pumped by a  
frequency-doubled
continuous-wave Nd:YAG laser (532 nm). 
The output of the OPA is analyzed by a  homodyne detector whose  
difference
current $i_-$ is directly proportional to the OPA's output electric  
field.
In order to obtain simultaneously information about the quantum   
states of 
all modes
emitted by the OPA, the homodyne detector current $i_-$ was recorded
with a bandwidth covering the whole frequency range up to 30 MHz,
exceeding the OPA's  17.5 MHz cavity linewidth. For this purpose
fast photodiodes (Epitaxx ETX500T, specified bandwidth 140 MHz)
with broadband amplifiers and a 30 MHz A/D-board (IMTEC) for
data collection were employed. 
The $i_-$ data (about 500\,000 points with 12 bit resolution) 
are taken
while the local oscillator phase is swept by $2\pi$ in  
approximately 8 ms.
The recorded noise trace is split 
into 16 separate traces, each containing the information of a 
mode pair of small bandwidth 1.9 MHz, centered at a sideband frequency 
$\omega\pm\Omega$. 
This is done by dividing the 
Fourier transform of the recorded trace into 16 equal intervals,  
and taking the
inverse transform of each interval separately. The same is done for a 
measured $i_-$-trace of a vacuum state, in order to
normalize each of the 16 traces. Due to electronic excess noise 
of the detection system at low frequencies 
the first of the 16 traces, containing the spectrum between 0 and  
1.9 MHz,
is discarded.  The quantum states of the remaining 15 modes can be 
reconstructed the same way as described in \cite{bre97a}, employing 
the inverse  Radon transform \cite{vog89} and the
pattern functions of the harmonic oscillator \cite{dar94}. 
Thus a whole  ``spectrum'' of Wigner functions and  
density matrices is obtained from a single temporal record.

\begin{figure}
\epsfxsize=15cm
\epsfbox[0 460 430 770]{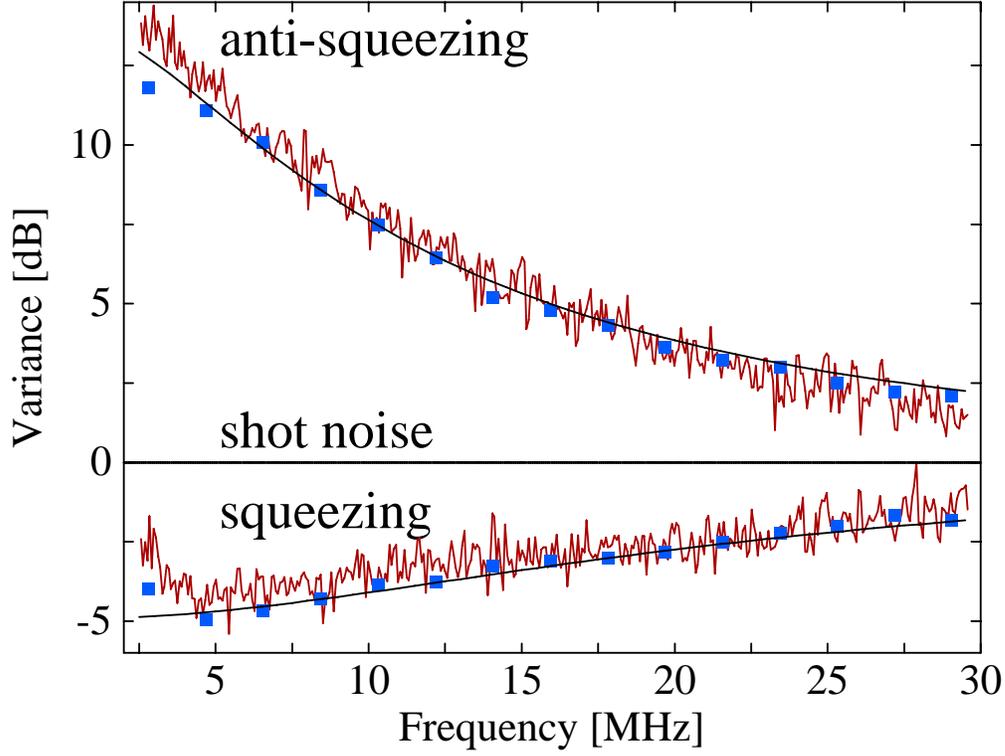}
\caption{
Spectrum of squeezing and anti-squeezing:
Spectrum analyzer traces (upper and lower grey lines)
in comparison with a theoretical fit (smooth black lines) using  
eq.1. Each of the 15 pairs of 
black squares corresponds to one reconstructed quantum state at a  
specific 
frequency $\Omega$ and represents the  maximum/minimum width of the
sampled marginal distributions.
The reduction in squeezing
at frequencies below 4.5 MHz is due to classical laser noise at low
frequencies.
\label{figmult2}
}
\end{figure}

The main difference between the squeezed vacuum states reconstructed at
different $\Omega$ is 
the amount of noise reduction (squeezing) and enhancement  
(antisqueezing).
The spectra of quantum noise power of the squeezed quadrature
$\Psi_-$ and of the antisqueezed quadrature $\Psi_+$
of an OPA on resonance are given by \cite{col85}
\begin{equation}
  \Psi_\pm(\Omega, P)  = \Psi_0\,\left(1 \pm  
\xi\eta\,\frac{4\,d}{(\Omega/\Gamma)^2 + (1 \mp d)^2}\right)\; ,
\label{eqsqspec}
\end{equation}
where $d=\sqrt{P/P_{th}}$ is the pump parameter with pump  power $P$
and a threshold power $P_{th}$ of the OPA,
$\Gamma/2\pi=17.5$ MHz is the cavity linewidth (HWHM),
$\Psi_0$ is the spectral density of the vacuum state,
$\xi=0.88$ is the escape efficiency of the resonator, and
$\eta$ is the detection efficiency. The latter includes propagation  
losses
after the OPA, homodyne efficiency and detector quantum efficiency. 
In contrast to our previous work
we did not employ a mode cleaning cavity, thus classical 
noise of the pump wave was not negligible
and the modematching between local oscillator and the
OPA output signal was $\approx 95\%$. Furthermore the efficiency
of the photo detectors had degraded slightly from the 
previously measured 95-97\%, thus the overall detection
efficiency  including escape efficiency $\xi\cdot\eta$ 
was  fitted to be 0.7. The pump power was 
approximately half the threshold power.
Fig.\ref{figmult2} shows the spectra 
of the maximally squeezed and anti-squeezed quadratures
measured directly with
a spectrum analyzer in comparison with the ones 
obtained via multimode reconstruction.

\begin{figure}
\epsfxsize=16cm
\epsfbox[40 260 470 580]{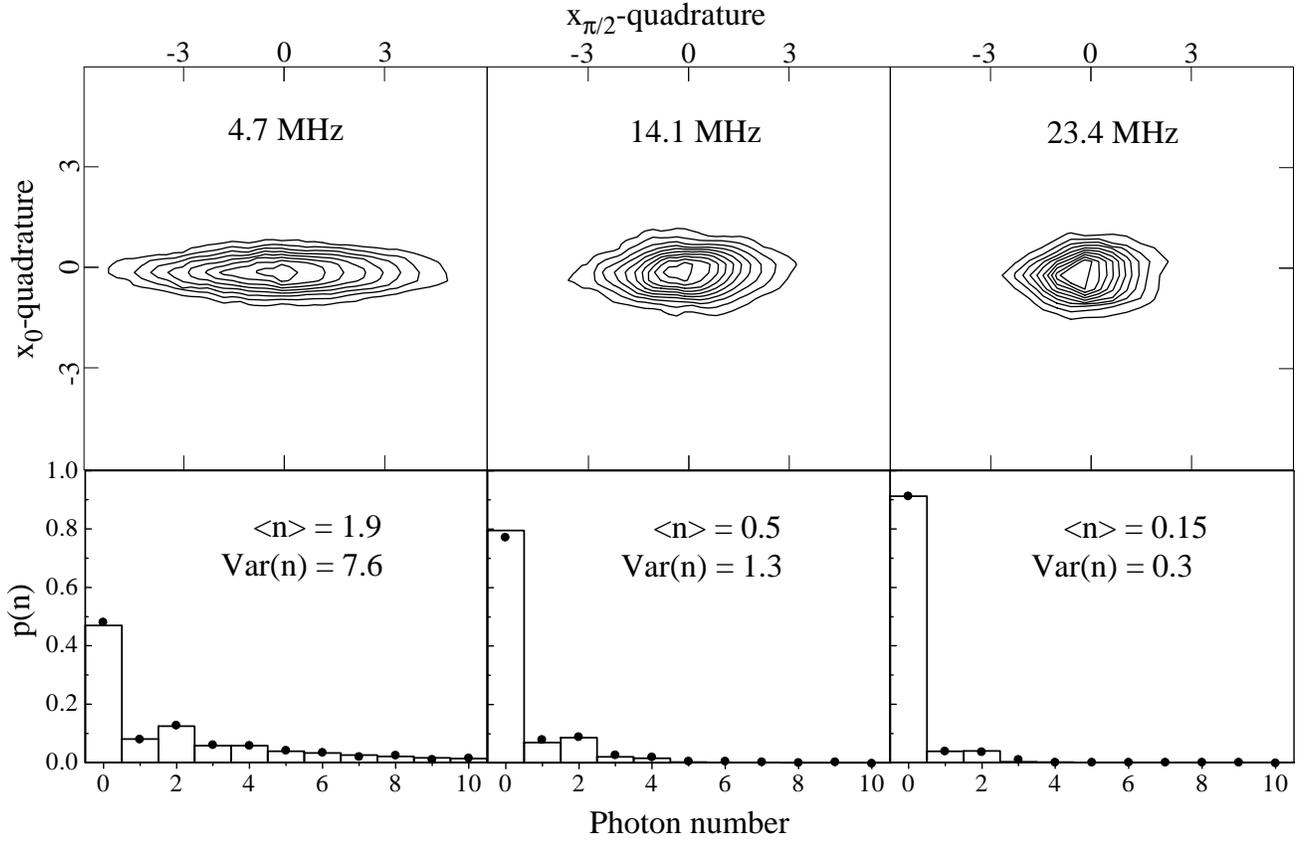}
\caption{
Three examples of the simultaneously recorded  squeezed vacuum  
states emitted by the OPA:
Upper row: contours of the Wigner functions, lower row: corresponding
reconstructed photon number distributions (dots) with theoretical 
expectations (histograms). For increasing offset  frequency $\Omega$ 
from the OPA cavity center frequency
the states aproach the vacuum state and the characteristics of
squeezing such as photon number oscillations and ellipticity of the
Wigner function vanish.
\label{figmult3}}
\end{figure}

Three of the reconstructed states are plotted in Fig.\ref{figmult3}. 
They display strongly elliptical
Wigner functions and oscillations in the photon number \cite{bre95a}, 
characteristics that vanish when the state 
aproaches a 
vacuum state at frequencies sufficiently away from the OPA cavity  
center frequency. 
We remark that we did not include error corrections in the  
reconstruction
algorithm, as described in
\cite{dar95}.  
For a single mode treatment of previously measured data of us
that takes into account the finite detection efficiency see  
\cite{tan97}.
Note also that the present measurement method does not allow to  
make any statements about
possible correlations between the different mode-pairs. 
Regarding the light field generated by the OPA this is of minor  
importance, since
according to theory mode-pairs at different frequencies should be 
uncorrelated \cite{col85}.

\begin{figure}
\epsfxsize=9cm
\epsfbox[50 300 290 620]{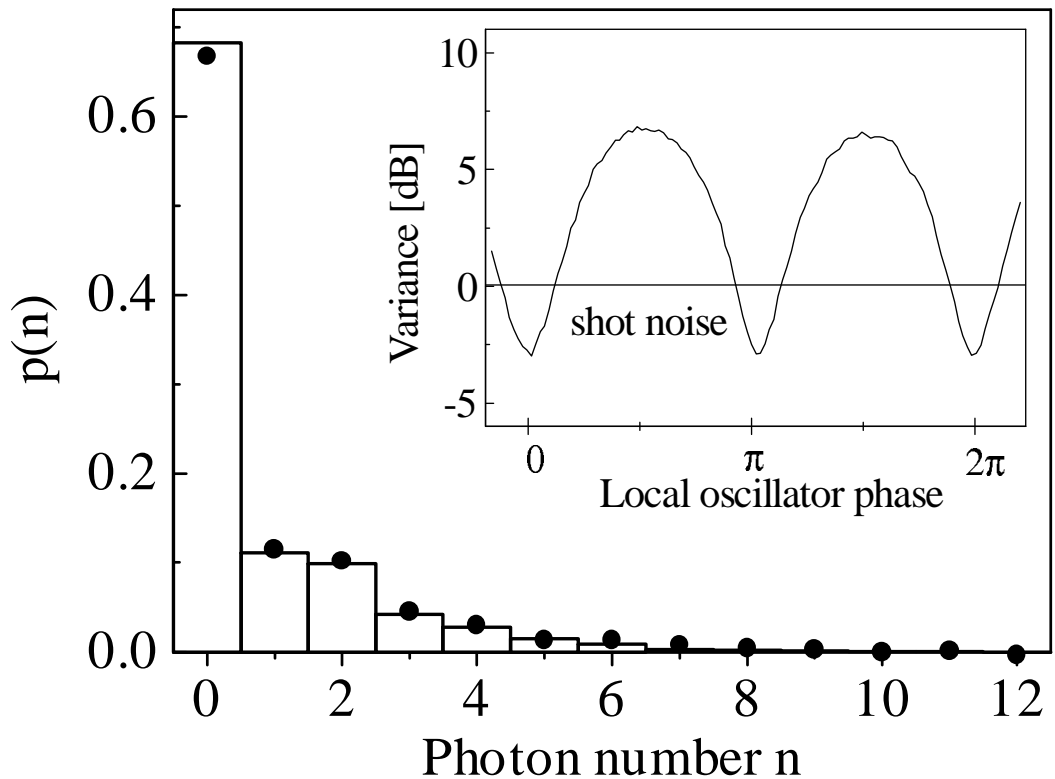}
\caption{
Electric field variances in dependence of the local oscillator phase
and photon statistics of  the sum of all OPA modes between 2 and 30 MHz.
The average photon number $\langle n \rangle$ = 0.8
gives the average photon flux per Hz bandwidth. 
This
implies  a total photon flux of  $2.2\cdot10^8$ photons/s $\approx  
45$ pW
within the detection bandwidth of 28 MHz.
\label{figmult4}
}
\end{figure}

A direct measure of the total noise of the 
squeezed and anti-squeezed quadratures 
$\int\Psi_\pm(\Omega, P)d\Omega$
of the overall OPA output within the bandwidth
1.9-30 MHz is obtained by analyzing the full quantum noise  
instead of its particular Fourier components.
(For this, the non-uniform spectral response of the A/D-board has  
to be eliminated
via normalization in Fourier space.)
The inset of Fig.\ref{figmult4} shows 
the total electric field variances 
as a function  of the local oscillator phase.
Well-known figures of this type usually depicted only the $E$-field
variances of a single mode of the light field
\cite{squeez}. 
Here we obtain for the overall OPA output a total squeezing of  -2.9 dB 
(= 0.47 linear scale) and 6.7 dB (= 4.7) for the total antisqueezing. 
The corresponding photon statistics is shown as well.
This multimode light field would be detected if one would employ
photon counters instead of homodyne detection.
Note that no photon number oscillations are observable in the total  
OPA output.
This may seem surprising, since almost all the photon number  
distributions
of the individual modes show oscillations and all the reconstruction
transformations are linear.
However, according to
basic laws of probability theory the overall sampled distribution of the
quantum noise is not given by the average but by the convolution of 
the distributions of the individual modes. Therefore 
the photon statistics of the 
total OPA output is not given by averaging the single mode statistics.
What is averaged though in our case of uncorrelated Gaussian noise  
distributions
is, as mentioned above, the variance at each particular phase angle.

A further perspective is gained by the 
analysis of our data in the time domain. 
At a fixed phase $\theta$ of the local oscillator, 
the recorded time trace can be 
regarded as the quantum trajectory of the quadrature 
$x_\theta(t) = 2^{-1/2}(a^\dagger(t)\,e^{{\rm i}\theta}\, +  
\,a(t)\,e^{-{\rm i}\theta})$, 
where $a(t)$ is the Heisenberg output field operator.
Thus the recorded data contain 
all information needed to extract the first order
correlation function. This function
is easily calculated using the input-output formalism for optical
cavities \cite{col85}.
\begin{eqnarray}
g_{(1)}(\tau) & \equiv & 
  \frac{ \langle a^{\dagger}(\tau) a(0) \rangle }
   {\langle a^{\dagger}(0)a(0)\rangle } \nonumber \\
& & \nonumber \\
&  =   & \frac{ \langle x_\theta(\tau) x_\theta(0) \rangle_\theta }
   {\langle x_\theta(0)x_\theta(0)\rangle_\theta }\\
& & \nonumber \\
& = & \frac{1-d^{2}}{2d} \left[ \frac{1}{1-d}
e^{-(1-d)\Gamma\tau} - \frac{1}{1+d}e^{-(1+d)\Gamma\tau}
\right] \,  \nonumber .
\end{eqnarray}
\noindent 

\begin{figure}
\epsfxsize=16cm
\epsfbox[40 480 490 770]{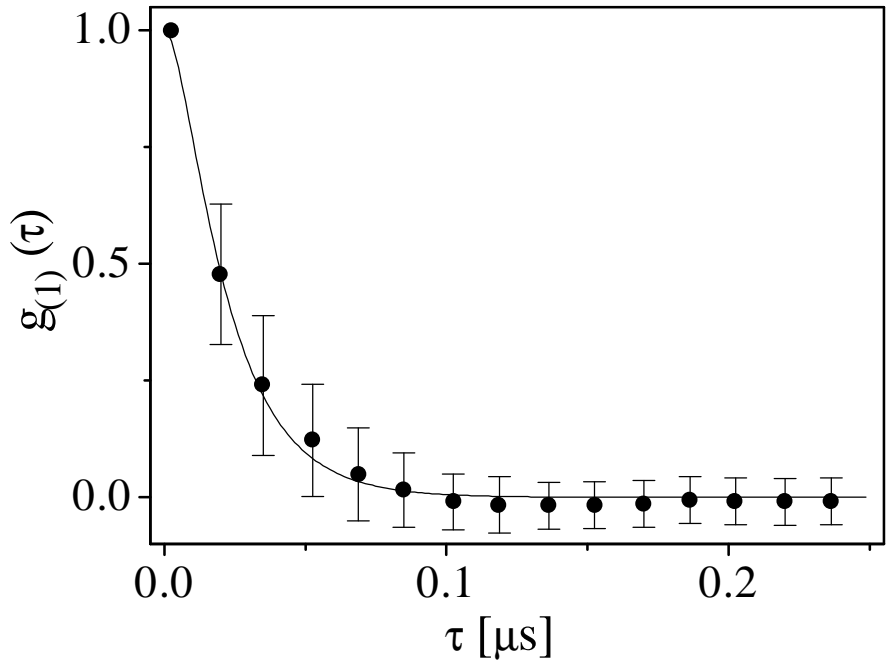}
\caption{
First--order time correlation function of the  
recorded noise trace (dots)
in comparison with theory (line). A quasi-exponential decay similar  
to the one of a thermal state is observed.
\label{figg1}
}
\end{figure}

Here  $<>_\theta$ means averaging over all phase angles.
Fig.\ref{figg1} shows that the evaluation of the first order 
correlation function obtained from the experimental data is in good 
agreement with theory. 
In principle it is possible to obtain via broadband homodyne detection
all higher order time correlations 
of the field quadratures 
$\langle x_\theta(\tau_1)x_\theta(\tau_2) ... x_\theta(\tau_n)\rangle$.
Except for one special case \cite{man95} their significance does  
not appear to 
have been studied.
For our data the evaluation of these correlations could not be done  
reliably,
due to difficulties with a proper normalization at $\tau=0$ and
since the minimum recordable timestep of 1.7 ns was too large
in comparison to the expected rapid exponential decay times  
$e^{-n(1-d)\Gamma\tau}$.

Note also that since a homodyne measurement is always a two-mode  
measurement
of the pair of modes at frequencies $\omega\pm\Omega$,
our experiment shows two-mode time correlations. For single-mode
time correlation measurements a heterodyne system (detection  
efficiency $\eta<$0.5)
can be employed. This is possible,
since detection efficiency does not play the same 
crucial role in time correlation 
measurements that it does for quantum state reconstructions, in  
fact ideally
$g_{(n)}$ is independent of $\eta$.

In conclusion, we have demonstrated the first simultaneous  
reconstruction of a whole
spectrum of quantum states of the optical light field. 
Some of the reconstructed states emitted from the OPA
display photon number oscillations and  strongly elliptical
Wigner functions. These characteristics vanish for the mode-pairs
at frequencies sufficiently away from the OPA cavity center frequency. 
Furthermore the first order time correlation 
function was evaluated in good agreement with theory.

Our measurement scheme may be also useful for a variety of other
quantum optical systems with more complex frequency or time  
dependencies.
Examples are recent squeezing experiments with solitons in a fibre  
\cite{fri96},
pump-noise-suppressed laser diodes \cite{bec98},
or exciton-polariton systems in semi-conductors \cite{jia98,ray999}.

We thank Robert Bruckmeier and 
Kazimierz Rz\c{a}\.{z}ewski for very helpful discussions. 
Financial support  was provided by the Deutsche Forschungsgemeinschaft 
and the Esprit  LTR project 20029-ACQUIRE.
One of us (F.I.) also gratefully acknowledges financial support from the  
Alexander von Humboldt Foundation, and the hospitality of the LS
Mlynek at the Fakult\"at f\"ur Physik, Universit\"at Konstanz, while
on leave of absence from the Dipartimento di Fisica, Universit\`a di
Salerno.


\vspace*{2cm}


\newpage

\end{document}